\documentclass[11pt]{article}

\usepackage{alltt}
\usepackage{graphicx}
\usepackage[dvips]{epsfig}
\usepackage{amssymb}
\newcommand{\grad}{\nabla}
\newcommand{\half}{{\textstyle{\frac{1}{2}}}}
\newcommand{\Ref}[1]{\mbox{\rm{(\ref{#1})}}}

\newcommand{\lt}{<}

\usepackage{art11mod}
\usepackage{float}
\usepackage{numinsec}
\usepackage{bar}
\usepackage{url}

\newcommand{\R}{\mbox{${\mathbb R}$}}

\newfloat{Algorithm}{thp}{loa}[section]

\newcommand{\cA} {\mbox{$\cal A$}}
\newcommand{\cB} {\mbox{$\cal B$}}
\newcommand{\cD} {\mbox{$\cal D$}}

\begin{document}
\setlength{\baselineskip}{15pt}

\pagestyle {plain}

\thispagestyle{empty}
\vspace{1.75in}

\begin{center}

ARGONNE NATIONAL LABORATORY

9700 South Cass Avenue

Argonne, Illinois  60439

\vspace{1.5in}

{\large
{\bf 
GPCG: A CASE STUDY IN THE PERFORMANCE AND SCALABILITY OF
OPTIMIZATION ALGORITHMS
}
}

\vspace{.5in}

{\bf Steven J. Benson,
Lois Curfman McInnes, and
Jorge J. Mor\'e}

\vspace{.5in}

Mathematics and Computer Science Division

\vspace{.25in}

Preprint ANL/MCS-P768-0799

\vspace{.5in}

September 2000
\end{center}

\vspace{2.5in}

\bigskip

\par\noindent
This work was supported by the Mathematical, Information, and
Computational Sciences Division subprogram of the Office of Advanced
Scientific Computing, U.S. Department of Energy, under Contract
W-31-109-Eng-38.
\newpage
\mbox{}
\pagenumbering{roman}
\setcounter{page}{2}
\newpage
\pagenumbering{arabic}
\setcounter{page}{1}

\setcounter{page}{1}
\begin{center}

{\large \bf GPCG: A Case Study in the Performance and Scalability of
Optimization Algorithms\footnotemark}

\bigskip

{\bf Steven J. Benson,
Lois Curfman McInnes, and
Jorge J. Mor\'e}

\bigskip

Mathematics and Computer Science Division

Argonne National Laboratory

\end{center}

\footnotetext
{ This work was supported by the Mathematical, Information, and
Computational Sciences Division subprogram of the Office of Advanced
Scientific Computing, U.S. Department of Energy, under Contract
W-31-109-Eng-38.}

\begin{abstract}

GPCG is an algorithm within the Toolkit for Advanced Optimization (TAO)
for solving bound constrained, convex quadratic problems.
Originally developed by Mor\'e and Toraldo \cite{more-toraldo}, this algorithm
was designed for large-scale problems but had been implemented
only for a single processor.  The TAO implementation
is available for a wide range of high-performance architecture,
and has been tested on up to 64 processors
to solve problems with over 2.5 million variables.
\end{abstract}


\section{Introduction}

The Toolkit for Advanced Optimization (TAO)
focuses on the design and implementation of
component-based optimization software for the
solution of large-scale optimization applications.
Our approach is motivated by the
scattered support for parallel computations and
lack of reuse of linear algebra software in
currently available optimization software.
We exploit numerical abstractions in the optimization software
design so that we can leverage external parallel computing
infrastructure (for example, communication libraries and visualization
packages) and linear algebra tools in the
development of optimization algorithms. The algorithms in the toolkit
place strong emphasis on the reuse of external tools where appropriate.
Our design enables connection to lower-level
support (parallel sparse matrix data
structures, preconditioners, solvers) provided in toolkits such as
PETSc \cite{petsc,PETSc-user-ref},
and thus we are able to build on top of these toolkits
instead of having to redevelop code. The advantages in
terms of development time are significant.

Initial work in the TAO project \cite{tao-web-page,tao-user-ref}
has centered on the development of a
core library of components for various types of optimization problems,
including unconstrained and bound-constrained minimization and
nonlinear least squares.  To explain the TAO design strategy and
analyze parallel performance issues, we focus on the
gradient projection conjugate gradient (GPCG) algorithm for the solution
of the bound-constrained quadratic programming problem
\begin{equation} \label{def_bqp}
\min \{ q(x) : l \leq x \leq u \} ,
\end{equation}
where $q : \R^n \mapsto \R $ is a strictly convex quadratic function,
and the vectors $l$ and $u$ define bounds on the variables.
Although GPCG had been originally designed \cite{more-toraldo} 
for large-scale problems, implementation of GPCG on a 
parallel architecture presented significant obstacles that
are typical of a large class of optimization algorithms.
The most significant obstacle arises from the 
method used to compute the step between iterates.
Specifically, in modern active set methods for solving
\Ref{def_bqp}, the step between iterates is usually defined
via the approximate solution of a linear system of the form
\[
A_k w_k = - r_k ,
\]
where the matrix $ A_k $ and the vector $r_k$ are, respectively,
the reduced Hessian matrix and the reduced gradient
of $q$ with respect to the free variables.
In a parallel environment, the efficient implementation of the conjugate
gradient method requires that $A_k$ 
be evenly distributed over the processors,
but since the set of free variables can change
drastically between iterates, the reduced matrix is
unlikely to be well distributed. Hence,
a redistribution of the rows of $A_k$ over the processors
may be necessary at each iteration.

This observation implies that the
scalability of the GPCG is limited not only by the efficiency
of the redistribution algorithm but by the sizes of
the matrices $A_k$. If the set of free variables is large,
then performance is likely to improve because the
communication overhead is small, while performance is
likely to suffer when there are few free variables.
Thus, the GPCG algorithm is prime candidate for a case study
in the performance and scalability of optimization algorithms
in parallel architectures.

Our implementation of GPCG uses object-oriented techniques
to leverage the parallel
computing and linear algebra infrastructure offered by PETSc
\cite{petsc,PETSc-user-ref}, which relies on MPI \cite{using-mpi} for all
interprocessor communication.  
As a result, our implementation runs on a wide variety of
high-performance architectures.
Biros and Ghattas \cite{GB99b,GB99a} have been using
a similar approach for the solution of PDE-constrained optimization problems.
They have also been concerned with efficiency and scalability
issues, but for quadratic problems with linear equality constraints.
As we have pointed out, inequality constrained optimization
problems give rise to different performance issues.
Hohmann \cite{hohmann:94}, 
Deng, Gouveia and Scales \cite{HLD94},
Meza \cite{meza:94},
Bruhwiler et al. \cite{bsca98}, and Gockenbach, Petro, and Symes
\cite{Gockenbach:1999:CCL}
have employed object-oriented design
for nonlinear optimization, but their work
does not address the reuse of linear algebra toolkits and
is restricted to uniprocessor environments.
Our use of object-oriented techniques and linear algebra
toolkits also distinguishes our implementation of GPCG from
the data-parallel implementation of McKenna, Mesirov, and
Zenios \cite{MOM95}. In particular, they can rely only
on diagonal preconditioners, while our approach allows a wide
range of preconditioners.

Sections \ref{qp} and \ref{alg} are dedicated to
background material on the bound-constrained 
optimization problem \Ref{def_bqp} and to
a brief overview of the GPCG algorithm, while
Section \ref{design} has a discussion
of our design philosophy and its benefits in
developing robust and scalable solutions strategies.

The performance results
in Section \ref{sec:performance} are noteworthy in several ways.
First, the number of faces visited by GPCG is remarkably small.
Other strategies can lead to a large number of gradient projection
iterates, but the GPCG algorithm is remarkably efficient.
Another interesting aspect is that because of the
low memory requirements of iterative solvers, we are able
to solve problems with over 2.5 million variables
with only $ 8 $ processors.
Strategies that rely on direct solvers are likely to need
significantly more storage, and thus more processors.
Finally, these results show that the GPCG implementation has
excellent efficiency. 

Section \ref{sec:analysis} examines the scalability of the
GPCG component functions and the performance of GPCG
as the number of variables and the number of active
variables at the solution change. These results illustrate the
complex performance behavior for constrained optimization
problems as well as the observation that
performance results that focus only on efficiency can be
deceiving if the total computing time is not taken into account.

Section \ref{sec:preconditioners} considers the performance
of GPCG as the preconditioners change. The
ability to use various preconditioners is
a result of our design, which allows the
connection to external linear algebra toolkits.
Our results in this section show that for our benchmark
problem, a block Jacobi preconditioner with
one block per processor, where each subproblem is solved
with a standard, sparse ILU(2) factorization, is faster than
a variant with ILU(0). We also show that both block Jacobi variants are faster
than a simple point Jacobi method,
although the point Jacobi preconditioner exhibits
better scalability.

\section{Bound-Constrained Quadratic Optimization Problem}

\label{qp}

A classical result shows that the bound-constrained 
quadratic optimization problem (\ref{def_bqp}) has a  unique solution
on the feasible region
\begin{equation} \label{def_bounds}
\Omega = \{ x \in \R^n: l \leq x \leq u \}
\end{equation}
when the quadratic $q : \R^n \mapsto R $ is strictly convex, so that
\begin{equation} \label{def-quadratic}
q(x) = \frac{1}{2}x^TAx + b^Tx +c,
\end{equation}
where $A \in \R^{n \times n}$ is symmetric and
positive definite, $ b \in \R^n$, and $c \in \R$.
This result holds for unbounded $\Omega$, and we thus
allow the components of $l$ and $u$ to be infinite.
Solutions to problem \Ref{def_bqp} satisfy the Kuhn-Tucker conditions
\[ \begin{array}{lllll}
\partial_iq(x) & = & 0 & \mbox{ if } & x_i \in (l_i, u_i) \\
\partial_iq(x) & \geq & 0 & \mbox{ if } & x_i = l_i \\
\partial_iq(x) & \leq & 0 & \mbox{ if } & x_i = u_i ,\\
\end{array}
\]
where $\partial_iq(x)$ is the partial derivative of $q$ with
respect to the $i$th variable.
Approximate solutions can be defined in terms of the projected
gradient, defined by
\begin{equation} \label{proj-gradient}
 \left[ \nabla _{\Omega} q(x) \right] _i = \left\{
\begin{array}{lll}
\partial_i q(x) & \mbox{ if } & x_i \in (l_i, u_i) \\
\min \{ \partial_i q(x),0 \} & \mbox{ if } & x_i = l_i \\
\max \{ \partial_i q(x),0 \} & \mbox{ if } & x_i = u_i
\end{array}
\right.
\end{equation}
This definition of a projected gradient is appropriate 
because $x^*$ is a solution of (\ref{def_bqp}) if and only if
$\nabla_{\Omega} q(x^*)=0$.

Given $x_0 \in \Omega$, and a tolerance $\tau$, an
approximate solution to the bound constrained problem (\ref{def_bqp}) 
is any vector $x \in \Omega$ such
that
\begin{equation} 
\label{bqp_approximate_sol}
\| \nabla_{\Omega} q(x) \| \leq \tau . 
\end{equation}
Note that (\ref{bqp_approximate_sol}) holds whenever
$x$ is sufficiently close to $x^*$ and in the face of $\Omega $ that
contains $x^*$.  The concept of a face is standard in convex analysis;
for the convex set (\ref{def_bounds}), the face of $\Omega$ that contains
$x$ is
\[ 
\Bigl \{ y \in \Omega: y_i = x_i \mbox{ if } x_i \in \{ l_i, u_i \} \Bigr \}. 
\]
Thus, the face of the feasible set that contains $x$ can
be described in terms of the set of active constraints
\[ 
{\cal A}(x) = \{ i: x_i = l_i \mbox{ or } x_i = u_i \}. 
\]
Variables with indices in ${\cal A}(x)$ are the active variables,
and those with indices outside  ${\cal A}(x)$ are the free variables.
Similarly, the binding variables are those with indices in
\[ 
{\cal B}(x) = \{ i:x_i=l_i \mbox{ and } \partial_iq(x) \geq 0,
\mbox{ or } x_i=u_i \mbox{ and } \partial_iq(x) \leq 0 \}. 
\]
The Kuhn-Tucker conditions show that 
$ \cB (x) = \cA (x) $ at a solution, so that if all the active
variables are not binding, then $x$ is not on the face
that contains the solution.

\section{The GPCG Algorithm}

\label{alg}

The GPCG algorithm uses a gradient projection method
to identify a face of the feasible region $ \Omega $
that contains the solution, and the conjugate gradient
method to search the face.
This section provides an outline of the algorithm and notes any
differences between our implementation and the implementation
of Mor\'e and Toraldo \cite{more-toraldo}.

Given $y_0=x_k$,
the gradient projection method generates a sequence of vectors $\{y_j\}$
in the feasible region $\Omega$ such that
\begin{equation} \label{next-y}
y_{j+1} = P [ y_j - \alpha_j \nabla q(y_j) ],  
\end{equation}
where $P$ is the projection
onto (\ref{def_bounds}), and
the step size $\alpha_j$ is chosen such that
\begin{equation}  \label{gplsstop}
 q(y_{j+1}) \leq q(y_j) + \mu
\langle \nabla q(y_j), P [ y_j - \alpha_j \nabla q(y_j) ] - y_j \rangle
\end{equation}
for some $\mu \in (0, 1/2 )$.
The projection $P$ can be computed in $n$ operations by
\[ 
P[x] = \mbox{ mid} (l,u,x), 
\]
where $\mbox{ mid} (l,u,x)$ is the vector whose $i${th} component
is the median of the set $\{ l_i, u_i, x_i \} $.
The step size is computed by a
projected search \cite{more-toraldo} by setting $ \alpha_j $
to the first member of the sequence
$ \alpha_0 ( \half ) ^ j $ for $ j = 0, 1, \ldots $ such that
$ y_{j+1} $ satisfies the sufficient decrease condition \Ref{gplsstop}.
In our implementation, we use
\begin{equation} \label{bqpls}
\alpha_{0}= \arg \min 
\left \{ 
q \left (y_k - \alpha  \nabla_{\Omega} q(y_k) \right ):
\alpha > 0 
\right \} .
\end{equation}
Computation of $ \alpha_0 $ is straightforward, since 
the mapping 
$ \alpha \mapsto 
q \left (y_k - \alpha  \nabla_{\Omega} q(y_k) \right ) $ is
a quadratic.

We generate gradient projection
iterates until sufficient progress is not made or
the active set settles down.
Thus, we generate iterates until either
\begin{equation} 
\label{pgstop1}
 {\cal A}(y_j) = {\cal A}(y_{j-1})
\end{equation}
or
\begin{equation} 
\label{pgstop2}
 q(y_{j-1}) - q(y_j) \leq \eta_1 \max \{q(y_{l-1}) - q(y_l) : 1 \leq l < j \}.
\end{equation}
If either test is satisfied, we proceed to the
conjugate gradient part of the algorithm.

The first test (\ref{pgstop1}) measures when the
active set settles down. For nondegenerate problems, 
(\ref{pgstop1}) holds in a neighborhood of the solution.
The gradient
projection could be followed until the optimal face is found, but
experience has shown that a large number of iterates may be required.
The second test (\ref{pgstop2}) measures when
the gradient projection method is not making sufficient progress.

Given an iterate $x_k$ and the active set  ${\cal A}(x_k)$,
the conjugate gradient method computes an approximate
minimizer to the subproblem
\begin{equation} \label{cg}
\min \{ q(x_k + d): d_i = 0, i \in {\cal A}(x_k) \}.
\end{equation}
This problem is unconstrained in the free variables.  Note that
if $x_k$ lies in the same face as the solution and $d_k$ solves
(\ref{cg}), then $x_k + d_k$ is the solution of (\ref{def_bqp}).

The conjugate gradient algorithm for solving
\Ref {cg} is implemented by expressing
this subproblem in terms of an equivalent subproblem in the
free variables.
If $ i_1, \ldots, i_{m_k} $
are the indices of the free variables, and 
the matrix $Z_k$ is defined as the matrix in $\R^{n \times m_k} $ whose
$j$th column is the $i_j$th column of the identity matrix in
$\R^{n \times n}$, then
subproblem \Ref{cg} is equivalent to the
unconstrained subproblem
\begin{equation}
\min \{ q _ k ( w ) : w \in \R ^ { m_k } \}  ,
\label{cg2}
\end{equation}
where
\[
q_k(w) \equiv q ( x _ k + Z _ k  w ) - q ( x_k ) =
\half \langle w , {A_k} w \rangle + \langle r_k , w \rangle  .
\]
The matrix $ A_k $ and the vector $ r_k $ are, respectively, the
reduced Hessian matrix of $q$ and reduced gradient of $q$ at $x_k$
with respect to the free variables.
If $A$ is the Hessian matrix of the quadratic $q$, then
\[
A_k = Z_k^T A Z_k , \qquad r_k = Z_k^T \nabla q(x_k)  .
\]
Also note that $A_k$ is the matrix obtained from
$A$ by taking those rows and columns whose indices correspond
to free variables;
similarly, $r_k$ is obtained from $\nabla q(x_k)$ by
taking the components whose indices correspond to free variables.

Given a starting point $ w_0 \in \R^{m_k} $, the conjugate gradient algorithm
generates a sequence of iterates $ w _ 0 , w_1 , \ldots $
that terminates at a solution
of subproblem \Ref {cg2} in at most $m_k$ iterations.
We use the conjugate gradient algorithm 
until it generates $ w _ j $ such that
\begin{equation}
q_k ( w_{j-1} ) - q_k ( w _ {j} ) \le \eta_2 \max \{
q_k ( w_{l-1} ) - q_k ( w _ {l} ) : 1 \le l \lt j \}
\label{cgstop1}
\end{equation}
for some tolerance $ \eta_2 > 0 $. 
The approximate 
solution of \Ref {cg} is then
$ d _ k = Z_k w _ {j_k} $, where $j_k$ is the first index $j$
that satisfies \Ref {cgstop1}.

The termination test \Ref{cgstop1} is not standard.
Iterative solvers usually terminate when
\[
\| r_j + A_j w_j \| \le \eta_2 \|r_j\| 
\]
for some tolerance $\eta_2 \in (0,1) $.
This test suffers from the erratic behavior
of the residual $ \| r_j + A_j w_j \| $.
On the other hand, the termination test \Ref{cgstop1}
depends on  whether the conjugate gradient method
is making sufficient progress.

Given the direction $ d_k $, we use a projected
search \cite{more-toraldo} to define 
$ x_{k+1} = P [ x_k + \alpha_k d_k] $, where
$ \alpha_k $ is the first element in
the sequence $ ( \half ) ^ k $ for $ k = 0, 1, \ldots $ such that
\begin{equation}  \label{cglsstop2}
 q( x_{k+1} ) \le  q(x_k) + \mu
\langle \nabla q(x_k), x_{k+1} - x_k \rangle .
\end{equation}
More sophisticated projected searches are possible \cite{more-toraldo}
,
but this simple search has proved to be sufficient in all cases tried.
If
\begin{equation}
\label{cgstop3}
\cB ( x_{k+1} ) = \cA ( x_{k+1} ) ,
\end{equation}
then we find a more
accurate solution to subproblem \Ref{cg2} by reducing $ \eta_2 $
and continuing with the conjugate gradient method.
Otherwise, we terminate this iteration.

\begin{Algorithm}
\noindent{\bf Algorithm GPCG}
\begin{list}{}
{
\setlength{\parsep}{0pt}
\setlength{\itemsep}{0pt}
\setlength{\topsep}{0pt}
}
\item[]
Choose $ x_0 \in \Omega $.
\item[]
For $ k = 0, \ldots, $
\begin{list}{$\bullet$}
{
}
\item[]
Set $y_0 = x_k$, and generate gradient projection iterates
$y_1, \ldots, y_{j_k}$, where $j_k$ is the first index to satisfy
(\ref{pgstop1}) or (\ref{pgstop2}). 
Set $x_k= y_{j_k}$.

\item[]
Set $ w_0 = 0 $, and
generate conjugate gradient iterates $ w_1 , \ldots, w_{j_k} $
for the reduced system (\ref{cg}).
Set $ d _ k = Z_k w _ {j_k} $, where $j_k$ is the first index
that satisfies \Ref {cgstop1}.
\item[]
Use a projected search to generate $ x_{k+1} $.
If \Ref{cgstop3} holds, reduce $\eta_2$, and
continue with the conjugate gradient method.
\end{list}
\end{list}
\end{Algorithm}

Our outline of algorithm GPCG does not include the termination test.
An advantage of the termination test \Ref{bqp_approximate_sol} is 
that this test is satisfied \cite{JVB92}
in a finite number of iterations.
On nondegenerate problems GPCG terminates \cite{more-toraldo}
at the solution in a finite number of iterations.

Algorithm GPCG is suitable for large problems.
As opposed to some other active set methods,
each iteration is capable of
adding or removing multiple constraints from the active set.
Moreover, as we shall see, GPCG tends to require few iterations
for convergence.
Another advantage of the GPCG algorithm is that convergence
can be achieved while requiring only approximate solutions
to the linear systems.

\section{Software Design} \label{design}

The TAO design philosophy uses object-oriented techniques
of data and state encapsulation, abstract classes, and
limited inheritance to create a flexible optimization toolkit.
This section provides a short introduction to our design
philosophy by describing the
objects needed to create GPCG.

Our current implementation leverages the parallel computing
and linear algebra infrastructure offered by PETSc \cite{petsc,PETSc-user-ref},
which employs MPI \cite{using-mpi} for all interprocessor communication.
TAO optimization algorithms use high-level abstract data objects that
are provided by PETSc, including vectors, matrices, and index sets.
In this context, a vector ({\texttt Vec}) is an abstraction of an
array of values that represent a discrete field, and a matrix
({\texttt Mat}) represents a discrete linear operator that maps
between vector spaces.  An index set ({\texttt IS}) is a
generalization of a set of integer indices, which can be used for
selecting, gathering, and scattering subsets of vector and matrix
elements.  TAO also interfaces to the linear solvers (\texttt {SLES}) within
PETSc.  Because each of these abstractions has several underlying 
representations, TAO has easy access to a variety of parallel vector
and sparse matrix implementations as well as preconditioners and
Krylov subspace methods.

Solving an optimization problem with TAO requires first creating a
context data type called \texttt{TAO\_SOLVER}, which encapsulates
information about the solution process, including the algorithm,
convergence tolerances, options, and parameters.  All of the
computations and communications related to a particular solution
process are managed in the solver context variable.  After defining
the optimization problem, the user then calls \texttt{TaoSolve} to
determine the solution.  Finally, the user destroys the TAO solver via
\texttt{TaoDestroy}.  The code fragment in Figure \ref{tao_interface}
shows the main functions needed to solve bound-constrained quadratic
programming problems with TAO.

\begin{figure}[htb]
\medskip
\begin{alltt}
  TaoCreate(MPI_Comm comm,TaoMethod method,TAO_SOLVER *tao); 
  TaoSetQuadraticFunction(TAO_SOLVER tao,Vec X,Vec G,Mat A,Vec B,double c);
  TaoSetVariableBounds(TAO_SOLVER tao,Vec XL,Vec XU);
  TaoSolve(TAO_SOLVER tao);
  TaoDestroy(TAO_SOLVER tao);
\end{alltt}
\caption{TAO interface for GPCG\label{tao_interface}}
\end{figure}

The function \texttt{TaoCreate} creates the \texttt{TAO\_SOLVER}
context for one of several possible methods (denoted by
\texttt{TaoMethod}) for solving the problem.  This interface serves
several algorithms for bound-constrained quadratic problems in
addition to GPCG, including limited memory variable metric, trust
region Newton, and interior point techniques.  Moreover, this single
interface serves other types of optimization problems as well.
The function 
\texttt{TaoSetQuadraticFunction} in Figure \ref{tao_interface}
defines the objective
function (\ref{def-quadratic}) in terms of the \texttt{Mat} object \texttt{A},
\texttt{Vec} object \texttt{B}, and scalar \texttt{c}
and provides the \texttt{Vec} objects \texttt{X} and \texttt{G} 
that are used for the solution and gradient.

The function
\texttt{TaoSetVariableBounds} defines upper and lower bounds for the variables
\texttt{X} with the \texttt{Vec} objects \texttt{XL} and \texttt{XU}.
Additional routines may be used to specify the starting point
and various options for the optimization solver,
but the structure in  Figure \ref{tao_interface} is needed in all cases.
Detailed information can be found
in the TAO User Guide \cite{tao-web-page,tao-user-ref}.

TAO implements the GPCG algorithm as a sequence of well-defined routines.
The evaluation of the function and gradient of
the quadratic $q$,
for instance, can be implemented through the standard
numerical operations of matrix-vector 
multiplication, vector inner product, and vector \texttt{saxpy}.
TAO passes \texttt{Mat} and \texttt{Vec} objects, whose
representation is independent of our implementation of GPCG, to 
external tools
that perform the numerical computations.
Additional work vectors required by the algorithm 
are created by calling a routine that
clones the variable vector \texttt{X} in Figure \ref{tao_interface}.

Users working in a parallel environment must provide TAO with data structures
\texttt{A}, \texttt{B}, \texttt{X}, \texttt{G}, \texttt{XL}, and \texttt{XU} 
that are properly distributed over the processors.  
Appropriate distribution allows efficient executions of the 
matrix-vector multiplication, vector inner product, and vector \texttt{saxpy}
operations.
Numerical toolkits such as PETSc facilitate the creation of these objects and
provide the functionality for most of the required numerical operations.


The operations required to implement the GPCG algorithm as outlined
in Section \ref{alg} include
the vector and matrix operations listed in the preceding paragraph,
functions to compute
the pointwise minimum and maximum of two vectors,
and a function that creates an
index set that defines the indices where the elements
of two vectors are equal.

At each iteration of the GPCG algorithm, we also need to apply
the conjugate gradient method to the matrix $ A_k $ corresponding
to the free variables.
This is an important phase of the computation because,
as we shall see in Section \ref{sec:performance}, at least $70\%$ of
the GPCG computing time is due to the conjugate gradient method.
An efficient parallel implementation of the conjugate gradient method
requires that the reduced matrix $A_k$ 
be evenly distributed over the processors,
but since the set of free variables may not be well distributed over the
processors, the reduced matrix may not well distributed---regardless of
how the matrix $A$ is distributed.
Since an unbalanced load can result in tremendous losses in
performance, a redistribution of the rows of $A_k$ over the processors
may be necessary.
We end this section by discussing the implementation of
the conjugate gradient method for solving the
reduced problem in the free variables.

At least two techniques exist for applying the conjugate gradient
method to the reduced system of equations.
One technique creates a second matrix $A_k$ that contains the
rows and columns of $A$ corresponding to the free variables,
and then applies the conjugate gradient method to the reduced system.
An alternative technique applies the conjugate gradient method
to the rows and columns of the full matrix $A$ specified by the
index set of the free variables.
In our implementation, we chose the first method.  Despite the
additional memory requirements and cost of copying data,
this method is simpler,
facilitates the preconditioning and load-balancing of the reduced matrix,
and was easily implemented with the utilities provided by PETSc.

Our implementation of GPCG calls 
\texttt{MatExtractSubmatrix(Mat,IS,IS,Mat *)}, which accepts the matrix
$A$ and the index set that identifies the set of free variables,
and creates the reduced matrix $A_k$.  
A call to \texttt{VecCreateSubVec(Vec,IS,Vec*)}
accepts the gradient vector and index set identifying the free variables
to create a new, reduced vector.
In a parallel environment, the index sets
also define the distribution of the reduced matrix over the 
processors.
These operations require a careful implementation when load
balancing issues are taken into consideration.

We interface to the
preconditioned conjugate gradient method provided by the
\texttt{SLES} component of PETSc.
We use the \texttt{SLES} object
to define this iterative method,
its preconditioner, the solution tolerance, and an initial point.  
The routine
\texttt{LinearSolve(SLES,Mat,Vec,Vec)} 
computes an approximate solution to the linear system
using the \texttt{SLES} object.
At each iterate we create the
conjugate gradient solver, apply it to the reduced linear
system, and then destroy it.

In the entire implementation of GPCG no assumptions are made about
the representations of data in the vectors and matrices.  This
approach eliminates some of the barriers in using independently
developed software components by accepting data that is independent of
representation and interfacing to numerical routines with the appropriate
data formats.

\section{Performance}

\label{sec:performance}

We have evaluated the performance of the GPCG implementation
on a variety of architectures.
The data presented in this section 
was generated on the IBM SP
(each processor has 256 MB RAM, 128 KB cache
for data, and a 32 KB cache for instructions)
at Argonne National Laboratory; performance trends
were similar on other machines.

\begin{figure} 
\centerline{\epsfig{figure = 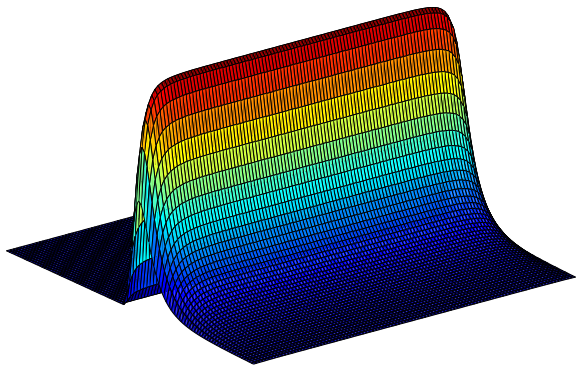, height=2.5in}}
\caption{The journal bearing problem with $ \varepsilon = 0.9 $.\label{pjb}}
\end{figure}

As a benchmark application we have used a journal bearing model,
a variational problem over a two-dimensional region.
This problem arises in the determination of the
pressure distribution in a thin film of lubricant
between two circular cylinders.
The infinite-dimensional version of this problem is
of the form
\[
\min \{ q ( v ) : v \ge 0 , \ v = 0 \mbox{ on } \partial D \} ,
\]
where $ v : \cD \mapsto \R $ is piecewise continuously differentiable,
$ q : H^1 \to \R $ is the quadratic 
\[
q (v) =
 \int_{ \cD } \left \{ \half  w_q (x) \| \grad v (x) \| ^ 2  -
  w_l (x) v (x) \right \} \, d x , 
\]
$ \cD = ( 0 , 2 \pi ) \times ( 0 , 2b ) $ for some constant $ b > 0 $,
and
\[
 w_q ( \xi_1 , \xi_2 ) = ( 1 + \varepsilon \cos \xi_1 ) ^ 3 , \quad
 w_l ( \xi_1 , \xi_2 ) = \varepsilon \sin \xi_1 ,
\]
where $ \varepsilon $ in $ (0,1) $ is the eccentricity parameter.
The eccentricity parameter
influences, in particular, the difficulty of the problem.
Figure \ref{pjb} shows the solution of the journal bearing problem
for $ \varepsilon = 0.9 $. The steep gradient in the solution
makes this problem a difficult benchmark.

Discretization of the journal bearing problem with either finite differences or
finite elements leads to a problem of the form \Ref{def_bqp} with
$ l \equiv 0 $ and $ u \equiv + \infty $.
The number of variables is $ n = n_x n_y $, where
$ n_x $ and $ n_y $ are, respectively, the number of
grid points in each coordinate direction of the domain $ \cD $.
See \cite{more-toraldo} for a description of the
finite element discretization.

We now analyze the performance of GPCG on large problems,
that is, problems that will not fit into the memory of a
single processor. Specifically,
we used a grid with 
$1600$ points in each direction, leading to a problem with
$ n = 2.56 \cdot 10^6 $ variables.

The initial point $ x_0 $ was set to the lower bound $l$.
We used $\eta_1  = 0.1$ in the test \Ref{pgstop2} to terminate the gradient
projection algorithm and $\eta_2=0.05$ in the test
\Ref{cgstop1} to terminate the conjugate gradient algorithm.
We stopped GPCG when the convergence test 
\Ref{bqp_approximate_sol} was satisfied with $ \tau = 10^{-4} $.

Table \ref{flops-2560K} presents performance data for
GPCG.
We show the number
of processors $p$,
the number of GPCG iterates (iters), the number of
conjugate gradient iterations $ n_{GP} $,
the wall clock solution time (in seconds),
the percentage of time ($t_{CG}$\%) used by the conjugate
gradient algorithm,
and the efficiency ($ \cal E $) of GPCG in going from 8 to 64 processors.
The time in the conjugate gradient algorithm
includes the time spent
computing the preconditioner.
Our design allows the use of several preconditioners, but for the
results in this section we used a block Jacobi preconditioner
with one block per processor, where each subproblem was solved
with ILU(2).

\begin{table}[htbp]
\caption{Performance of GPCG on the journal bearing problem
with $ n = 2.56 \cdot 10^6 $.}
\label{flops-2560K}
\begin{center}
\footnotesize
\begin{tabular}{| c r | c c r c r |}
\hline
\multicolumn{1}{|c}{$ \varepsilon $} & 
\multicolumn{1}{c|}{$ p $} & 
\multicolumn{1}{c}{iters} &
\multicolumn{1}{c}{$n_{GP}$} & 
\multicolumn{1}{c}{time} &
\multicolumn{1}{c}{$t_{CG}$\%} & 
\multicolumn{1}{c|}{$ \cal E $} \\ \hline
0.1  & 8 & 46 & 431 & 7419 & 86 & 100  \\ 
0.1  & 16 & 45 & 423 & 3706 & 83 & 100  \\
0.1  & 32 & 45 & 427 & 2045 & 82 & 91 \\
0.1  & 64 & 45 & 427 & 1279 & 82 & 73 \\
\hline
0.9  & 8 & 37 & 105 & 2134 & 70 & 100 \\
0.9  & 16 & 37 & 103 & 1124 & 71 & 95 \\
0.9  & 32 & 38 & 100 & 618 & 69 & 86 \\
0.9  & 64 & 38 & 99 & 397 & 68 & 67 \\
\hline
\end{tabular}
\end{center}
\end{table}

The results in Table \ref{flops-2560K} are noteworthy is several ways.
First, the number of iterations of GPCG is remarkably small.
This is surprising because the feasible set \Ref{def_bounds}
has $ 3^n $ faces, and the
GPCG visits only one face on each iteration.
Other strategies can lead to a large number of 
iterates, but the GPCG algorithm is remarkably efficient.

Another interesting aspect of the results in Table \ref{flops-2560K}
is that due to the
low memory requirements of iterative solvers, we were able
to solve these problems with only $ p = 8 $ processors.
Strategies that rely on direct solvers are likely to need
significantly more storage, and thus more processors.
Finally, these results show that the GPCG implementation has
excellent efficiency with 
respect to $ p = 8 $ processors,
ranging between $ 67\% $ and $ 100\% $.
This sustained efficiency is remarkable because the 
GPCG algorithm is solving a sequence
of linear problems with a
coefficient matrix set to the submatrix of the Hessian of $q$ with 
respect to the
free variables for the current iterate.
Thus, our implementation's repartitioning of submatrices deals effectively
with the load-balancing problem that is inherent
in the GPCG algorithm.

For these results we have noted that as $ \varepsilon $ increases,
both $t_{CG}\%$ and the overall efficiency decrease.
This observation follows from the empirical result
that the number of free constraints
at the solution is inversely proportional to the eccentricity
parameter $ \varepsilon $.
In particular, roughly $ 68 \% $ of the constraints are free at the
solution when $ \varepsilon = 0.1 $, and $ 54\% $ are free for
$ \varepsilon = 0.9 $.
Since the size of the linear system
that the conjugate gradient algorithm needs to solve increases
as $ \varepsilon $ decreases, the time required
by the conjugate gradient algorithm increases.
Since the parallel efficiency of larger problems is greater than the
parallel efficiency for smaller problems, the overall efficiency of GPCG
increases.

\section{Performance Analysis}

\label{sec:analysis}

GPCG is typical of optimization algorithms that
must deal with constrained problems in the sense that these
algorithms have dynamically changing active sets.
In this section we analyze the performance of GPCG.


Table \ref{flops-640K} presents performance results for the journal
bearing problem with dimension 640,000.
In comparing these results with those of the larger problem in Table
\ref{flops-2560K}, note that while the number of variables increases
by a factor of four,
the number of iterations, the number of gradient projection iterates,
and the time for solving the problem, increase by about a factor of two.
This seems to be fairly typical of GPCG but may not hold for
other optimization algorithms.
Some algorithms for unconstrained problems exhibit mesh invariance in the
sense that the number of iterations is independent of the number of
variables, but this does not generally hold for constrained problems.

\begin{table}[htbp]
\caption{Performance of GPCG on the journal bearing problem
with $ n = 640,000 $.}
\label{flops-640K}
\begin{center}
\footnotesize
\begin{tabular}{| c r | c c r c r |}
\hline
\multicolumn{1}{|c}{$ \varepsilon $} & 
\multicolumn{1}{c|}{$ p $} & 
\multicolumn{1}{c}{iters} &
\multicolumn{1}{c}{$n_{GP}$} & 
\multicolumn{1}{c}{time} &
\multicolumn{1}{c}{$t_{CG}$\%} & 
\multicolumn{1}{c|}{$ \cal E $} \\ \hline
0.1  & 2 & 27 & 227 & 2057 & 79 & 100  \\
0.1  & 4 & 26 & 227 & 1173 & 79 & 89 \\
0.1  & 8 & 27 & 232 & 639 & 78 & 80 \\
0.1  & 16 & 26 & 231 & 365 & 75 & 70 \\ 
0.1  & 32 & 27 & 230 & 220 & 74 & 58 \\
0.1  & 64 & 27 & 228 & 152 & 75 & 42 \\
\hline
0.9  & 2 & 21 & 58 & 645 & 65 & 100 \\
0.9  & 4 & 20 & 54 & 368 & 63 & 88 \\
0.9  & 8 & 20 & 52 & 199 & 64 & 81 \\
0.9  & 16 & 21 & 54 & 128 & 64 & 63 \\
0.9  & 32 & 20 & 52 & 74 & 61 & 54 \\
0.9  & 64 & 23 & 54 & 58 & 62 & 35 \\
\hline

\hline
\end{tabular}
\end{center}
\end{table}

When analyzing the parallel performance of an algorithm, we must bear
in mind that a problem can scale well only when the ratio of
computation to communication time is sufficiently large.  Thus, for a
particular problem size, scalability tapers off when more processors
are added than can be used effectively.  For GPCG, this effect can be
seen clearly by comparing the results in Table \ref{flops-640K} with
those in Table \ref{flops-2560K}.

An important aspect of the results in Table \ref{flops-640K} is that 
for this particular problem of dimension 640,000, the
efficiency of GPCG 
is acceptable for $ p \le 8 $ processors
but drops rapidly with more processors.
To explain the drop in efficiency, we
list in Table \ref{routines} the percentage of time spent in 
the main operations of GPCG.
Note that some of these operations overlap, so the sum
of the percentages always exceed $ 100\% $.
In this table {\it Vec Red} refers to
vector reductions, such as dot products and norms,
while {\it Vec Local} refers to vector operations such as
$  y \leftarrow \alpha x + y $.

\begin{table}[ht]
\caption{Scalability of GPCG functions
($ n=640,000 $, $ \varepsilon = 0.1 $) }
\label{routines}
\begin{center}
\footnotesize
\begin{tabular}{|c|ccccc|cc|}
\cline{2-8}
\multicolumn{1}{c|}{} &
\multicolumn{5}{c|}{Percentage of time}&
\multicolumn{2}{c|}{Total MFlops} \\
\hline
\multicolumn{1}{|c|}{Number}&
\multicolumn{1}{c}{Mat-Vec}&
\multicolumn{1}{c}{Vec} &
\multicolumn{1}{c}{Vec} &
\multicolumn{1}{c}{Linear}&
\multicolumn{1}{c}{Extract}&
\multicolumn{1}{|c}{Linear}&
\multicolumn{1}{c|}{TAO} \\

\multicolumn{1}{|c|}{Proc.}&
\multicolumn{1}{c}{Multiply}&
\multicolumn{1}{c}{Local} &
\multicolumn{1}{c}{Red} &
\multicolumn{1}{c}{Solve}&
\multicolumn{1}{c}{Submatrix}&
\multicolumn{1}{|c}{Solve}&
\multicolumn{1}{c|}{Solve} \\

\hline
1 & 27 & 15 & 7 & 81 & 1 &26 & 23 \\ 
2 & 30 & 15 & 8 & 83 & 2 &47 & 42 \\ 
4 & 30 & 12 & 8 & 82 & 2 &94 & 82 \\ 
8 & 29 & 11 & 10 & 81 & 2 &179 & 156 \\ 
16 & 26 & 10 & 14 & 78 & 2 &333 & 279 \\ 
32 & 24 & 9 & 22 & 78 & 2 &563 & 473 \\ 
64 & 20 & 5 & 36 & 78 & 2 &790 & 665 \\ 

\hline
\end{tabular}
\end{center}
\end{table}

The percentage of time spent in the various functions
of GPCG generally decreases slightly as the number of processors increases,
with the exception of the vector reductions.
Since vector reductions require
communication among all processors, they have a significant effect
on the efficiency of the algorithm.
Note that the time for vector reductions
remains fairly constant at about $8\%$ of the total computation
time for \mbox{1--8} processors but that the
efficiency of the algorithm declines
quickly as the percentage of time doing vector reductions
increases to $36\%$ on $64$ processors.
This analysis shows that 
the ratio of computation to communication for this problem is too small
for large number of processors and is responsible for
the loss in scalability of GPCG for $ p > 8 $.

In this discussion of efficiency bear in mind
that the Hessian matrix of the journal bearing problem is relatively
sparse with 5 nonzeros per row on average. 
The efficiency is likely to improve if we deal with
matrices with more nonzeros per row,
since then the amount of computation per conjugate gradient iteration 
increases.
These problems arise, for example,
in three-dimensional simulations or in variational
problems with vector functions, that is, variational
problems that require determining a vector-valued
$ v : \cD \mapsto \R^m $ for $ m > 1 $ that minimizes the quadratic $q$.

A surprising aspect of the results in Table \ref{flops-640K} is that
the percentage of time
required to extract the submatrix remains nearly constant
at $2\%$ of the total computation time, demonstrating the relative 
efficiency of this phase of the computation.
These results are surprising because at first sight the need to
extract an arbitrary submatrix and to refbalance the distribution of
rows across the processors would destroy the efficiency of the algorithm.
On the other hand, the creation of a second matrix to 
hold the submatrix requires
additional storage.  For large problems the additional storage
may exceed the memory capacity of a small number of processors.

Another important component of our scalability analysis is
the flop rate per processor.
As noted in Table \ref{flops-640K}, the flop rate for the
linear solve component of GPCG is 26 MFlops for
one processor and decreases to about $ 12.3 $ for 64 processors.
For comparison purposes, the flop rate of a Newton algorithm in
PETSc is about 42 MFlops for one processor on a system of nonlinear
equations with the same sparsity as the journal bearing problem.
This rate is higher than the rate achieved by the GPCG algorithm,
but this is to be expected because,  as previously mentioned, 
the GPCG algorithm spends a significant amount of time on tasks
with no arithmetic operations.  The extraction of the submatrix,
creating the reduced linear system and determining the
free variables, typically requires more than $ 10\% $ of the time.
Hence, it is unlikely that the GPCG algorithm, or any 
active set algorithm for constrained problems, can
achieve a computation rate as high as a Newton algorithm.

While these computations employed a standard 
compressed, sparse row format for matrix data,
higher flop rates could be obtained on
some problems, changing the matrix format.
Alternative storage
schemes that exploit the structured sparsity of these problems would
achieve higher flop rates for matrix operations by alleviating
unnecessary memory references.  Likewise, block sparse storage
variants for problems with multiple unknowns per grid point would
achieve higher flop rates \cite{gkmt98}.  Since our optimization
algorithms use a data-structure-neutral interface to matrix and vector
operations, we can easily experiment with such alternatives without
altering any of the optimization code.

\section{Preconditioners}

\label{sec:preconditioners}

The ability to experiment with various preconditioners is
a direct result of our design philosophy, which enables
connection to the 
linear algebra infrastructure provided in toolkits such as PETSc.
In particular, we compared the diagonal Jacobi
preconditioner with a block Jacobi preconditioner that used
one block per processor.
We employed sparse matrix based ILU as a subdomain solver for the block
Jacobi method, where we considered both ILU(0), which produced a
factored matrix that maintained the same sparsity pattern as the
subdomain matrix, and ILU(2), which allowed two levels of fill.

The statistics summarized in Table \ref{preconditioners}
are the eccentricity parameter $ \varepsilon$,
the number of processors $p$,
the number of iterations of GPCG,
the time required to solve the problem,
and the number of conjugate gradient iterations.
We present results only for $ n = 640,000 $, since similar
results were obtained for $ n = 2,560,000 $.

\begin{table}[htbp]
\caption{Performance of preconditioners in GPCG ($ n = 640,000 $)}
\label{preconditioners}
\begin{center}
\footnotesize
\begin{tabular}{| c r | c c c | c c c | c c c |}
\cline{3-11}
\multicolumn{2}{c}{} &
\multicolumn{3}{|c|}{Diagonal} &
\multicolumn{3}{c|}{Block Jacobi - ILU(0)} &
\multicolumn{3}{c|}{Block Jacobi - ILU(2)} \\
\hline
\multicolumn{1}{|c}{$ \varepsilon $} & 
\multicolumn{1}{c|}{$ p $} & 
\multicolumn{1}{c}{iters} &
\multicolumn{1}{c}{time} & 
\multicolumn{1}{c|}{CG iters} & 
\multicolumn{1}{c}{iters} &
\multicolumn{1}{c}{time} & 
\multicolumn{1}{c|}{CG iters} & 
\multicolumn{1}{c}{iters} &
\multicolumn{1}{c}{time} &
\multicolumn{1}{c|}{CG iters} \\ \hline
 0.1 &  4 & 26 & 2928 & 37045 & 27  & 1324 & 8679  & 26 & 1173 & 6312 \\
 0.1 & 16 & 26 & 851  & 37045 & 27  & 409  & 9105  & 26 & 364 & 6712 \\
\hline		                                   
 0.9 &  4 & 21 & 1216 & 18118 & 20 & 416  & 2654 & 20 & 368 & 1864\\
 0.9 & 16 & 22 & 390  & 18118 & 23 & 150  & 3390 & 21 & 128 & 2303\\

\hline
\end{tabular}
\end{center}
\end{table}

The number of GPCG iterations in Table \ref{preconditioners}
is independent of the number of
processors and of the preconditioner. 
In general we expect small variations in the number of iterations 
because different preconditioners create different approximate solutions
to linear systems and different paths to the solution.

In these experiments we were interested in the impact of
the preconditioner on the total time to solution. The Jacobi method is
scalable, so the main issue is whether the higher computational
cost of the block Jacobi is justified.
As expected, the block Jacobi preconditioner with subdomain solver ILU(2)
required fewer conjugate gradient iterations than subdomain solver ILU(0),
and both block Jacobi preconditioners required fewer
iterations than the point Jacobi method.
In addition, the block Jacobi methods also required less time.
In general, better preconditioners require more time to compute, and
this additional cost sometimes negates the savings achieved from fewer
iterations of the linear solver. 
In this problem, 
the block Jacobi preconditioners used
about half of  the time required by the diagonal preconditioner, and
the additional cost of computing better preconditioners is justified.
The most expensive preconditioner to compute of the three under consideration
in this work, namely, the block Jacobi method with subdomain solver ILU(2),
produced the fewest iterations by
the conjugate gradient method and the smallest overall solution time.

The ability to experiment easily with a variety of preconditioners is
an advantage because we can then choose a technique that is most
suitable to the problem.  In this spirit, we plan to experiment with
the evolving interfaces under development by the Equation Solver
Interface (ESI) \cite{esi-web-page} and Common Component Architecture
(CCA) \cite{cca99,cca-web-page} working groups, with a goal of
enabling dynamic use within TAO of any ESI-compliant preconditioning
components.

\section{Concluding Remarks}

We have shown that the TAO design leverages external
parallel computing infrastructure and
linear algebra toolkits to solve
large-scale optimization problems on high-performance
architectures.
With the exception of the work of 
Biros and Ghattas \cite{GB99b,GB99a},
other codes for large-scale optimization
problems are either custom-written
or restricted to uni-processor environments.

TAO \cite{tao-web-page,tao-user-ref}
extends to general nonlinearly bound-constrained optimization,
but the performance issues are more subtle due to the
impact of user-supplied function, gradient and Hessian code.
Extensions of TAO to large
linearly-constrained and nonlinearly-constrained optimization
problems is currently an active research area.

\section*{Acknowledgments}

The development of TAO would not have been possible without the
support and guidance of Satish Balay,
Bill Gropp, and Barry Smith.
They, together with Lois McInnes, are the main developers of PETSc.

\bibliographystyle{siam}

\bibliography{../tao,%
/home/more/papers/bibs/opt80,%
/home/more/papers/bibs/opt90}%

\end{document}